\documentclass[12pt,preprint]{aastex}

\usepackage{graphicx}
\def\beq{\begin{equation}}
\def\eeq{\end{equation}}
\def\beqar{\begin{eqnarray}}
\def\eeqar{\end{eqnarray}}

\def\fun#1#2{\lower3.6pt\vbox{\baselineskip0pt\lineskip.9pt
  \ialign{$\mathsurround=0pt#1\hfil##\hfil$\crcr#2\crcr\sim\crcr}}}

\def\qir{q_{\rm IR}}
\def\qir0{q_{\rm IR,0}}
\def\qfir{q_{\rm FIR}}

\usepackage{float}
\usepackage{lscape}

\begin{document}

\title{Impact of Galactic Interactions on the Evolution of the Far-Infrared-Radio Correlation}

\author{M. Pavlovi\'c}
\affil{Department of Physics, University of Novi Sad, Trg Dositeja Obradovi\'ca 4, 21000 Novi Sad, Serbia}
\affil{Department of Astronomy, University of Belgrade, Studentski Trg 16, 11000 Belgrade, Serbia}
\email{marina@df.uns.ac.rs}

\author{T. Prodanovi\'c}
\affil{Department of Physics, University of Novi Sad, Trg Dositeja Obradovi\'ca 4, 21000 Novi Sad, Serbia}
\email{prodanvc@df.uns.ac.rs}

\begin{abstract}

A strong correlation has been known to exist between the far-infrared (FIR) and radio emission of the star-forming galaxies. Observations have shown that although scatter is present, this correlation holds over a range of redshifts and does not evolve. However, there has been a number of more recent observations, especially in higher redshift surveys, indicating the opposite. The question that then presents itself is - what is driving this evolution? In this work we explore the possibility that the answer might be hiding in galactic interactions and revealed by morphology. We present a number of models based on the evolving number of galaxies of different morphological types, some of which  could potentially explain observed trends and scatter in general. 
Furthermore, we analyze a small sample of 34 submillimeter galaxies whose observations have been published and morphology classified. In this sample we look at the FIR-radio correlation separately in galaxies of different morphological types. We find that, while for both disk and irregular star-forming galaxies there are hints of evolution of this correlation with redshift, where this evolution appears to be stronger in irregular galaxies, due to low number statistics, both samples are also consistent with no evolution, making it at this point, difficult to discriminate between models. However, when analysis was performed on the combined sample, evolving and decreasing trend was indeed found, indicating, that evolution should be expected in at least one of the morphological types. 

\end{abstract}


\section{Introduction}
\label{sec:intro}

It has been known for a long time that there exists a strong correlation between the far infrared (FIR) and radio emission observed in star forming galaxies \citep{vanderkruit71, helou85, condon92, yun01}. The origin of this correlation has been attributed to essentially shared origin of emissions in these two bands, namely, young massive stars, which by being the source of the UV emission are responsible for heating up the dust that then emits infrared radiation, but which also, when they die as supernovae, are the source of the non-thermal radio emission coming from electrons accelerated in their remnants \citep{condon91}. It is thus clear that this correlation presents a powerful probe of the star-formation (SF) process \citep{condon92, bell03, murphy12}. Other commonly used SF tracers, such as the $\rm H_{\rm\alpha}$ or UV emission, have the disadvantage that they suffer from dust extinction and corrections for this process brings considerable uncertainty. The dust emission \citep{lisenfeld10}
in the mid-infrared (typically taken at 24 $\mu m$ or 70 $\mu m$), which is also frequently used as a SF
tracer on its own or in combination with $\rm H_{\rm\alpha}$ or UV, is not affected by dust extinction. However, these observations need to be done by space-satellite missions (Spitzer, Herschel, WISE) that have low spatial resolution. The radio emission avoids both of these disadvantages since it is not affected by the dust. A
second application of the FIR-radio correlation has been the use of the radio-submillimeter ratio as a photometric redshift indicator \citep{blain99}. The FIR-radio correlation has been used to identify and study radio-loud active galactic nuclei \citep{donley05, norris06, park08, delmoro13} and   to   estimate distances
and   temperatures   of   high-redshift   submillimeter   galaxies \citep{condon92, bell03, murphy12}. These estimates rely on the tightness of the FIR-radio correlation. The FIR-radio correlation is often quantified by the ratio parameter 
\begin{equation}
\label{eq:q}
q_{\rm FIR}=\log{\left( \frac{F_{\rm FIR}}{3.75 \times 10^{12} \rm W m^{-2}} \right) } - \log{\left( \frac{S_{1.4}}{\rm W m^{-2} Hz^{-1}} \right) }
\end{equation}
where $F_{\rm FIR}$ and $S_{1.4}$ are the rest frame far infrared emission flux density from $42 \rm \mu m$ to $122 \rm \mu m$ and the rest frame radio emission flux density at 1.4 GHz respectively \citep{helou85}. Although star-formation evolves over cosmic times, multiple studies have claimed that the FIR-radio correlation remains stable and that it does not evolve with redshift \citep{sargent2010}, but can instead be described with a single value. This value was determined from a sample of 1800 star-forming galaxies to be $q_{\rm FIR,0}=2.34 \pm 0.01$ \citep{yun01}. Nevertheless, there are models that point out that the evolution of this parameter, specifically its increase with redshift,  should, in fact, be expected due to increasing density of cosmic microwave photons with redshift, which would result in a more significant Inverse Compton energy loses, and consequently, decrease the non-thermal radio emission \citep{murphy2009}.  Recent observations of 12000 star-forming galaxies in the COSMOS field up to the redshift $z<6$ have revealed that this parameter, indeed evolves, but it however decreases with redshift as $q_{\rm FIR} (z)=(2.52 \pm 0.03)(1+z)^{-0.21 \pm 0.01}$ \citep{delhaize2017}. Therefore, either the radio emission is increasing or the infrared emission is decreasing more significantly with redshift. Some explanations for this observed trend can be found in a possible contamination by the presence of active galactic nuclei \citep{magnelli10, sajina08} or a larger contribution of thermal radio component \citep{delhaize2017}. One other possible explanation might come from increasing fraction of major mergers towards higher redshifts. Namely, major mergers of galaxies can cause enhanced synchrotron emission  through amplification of magnetic fields \citep{Kotarba2010}, through additional emission from gas bridges in taffy-like systems \citep{murphy13} or through acceleration of cosmic rays in tidal shocks that develop in the interstellar medium of interacting galaxies \citep{dp2015}.

In this paper we explore the possibility that major mergers are the underlying cause of the the decrease in the $q_{\rm FIR}$ parameter as reported in \citet{delhaize2017}. We do this by looking into the impact of morphology, as a proxy for interaction, on the evolution of the FIR-radio correlation with redshift.  Most galaxies in the local universe are either elliptical or disk galaxies, but there is also a small fraction of galaxies that do not resemble neither of the two - the irregular galaxies. It is widely accepted that irregular morphology is the indication of ongoing or past galactic interactions and that the number of irregular galaxies is increasing with redshift \citep{mortlock13, mundy17}. In Section 2 of this work we present  few interaction-based models that could possibly explain behavior observed by \citet{delhaize2017}. 
Following that, in Section 3 we analyze a small available sample of high-redshift submillimeter galaxies (SMG) observed in the COSMOS survey that have been morphologically classified \citep{miettinen2017a,miettinen2017b}, and test if the FIR-radio correlation changes with morphology in selected subsamples. 

\section{Evolution of the FIR-Radio Correlation}

Motivated by the observed decrease in the $\qfir$ found in \cite{delhaize2017} and the works of \cite{lisenfeld10} and \cite{dp2015}, here we model the evolution of $\qfir$ with redshift, by assuming that the $\qfir$ parameter found in irregular and interacting star-forming galaxies is different from the value that was well-established for star-forming galaxies at low-redshift. The deviation from the nominal value of $q_{\rm FIR,0}=2.34$ \citep{yun01}, and a greater scatter in general, will then follow if the observed sample of galaxies contains an increasing fraction of interacting and irregular systems towards higher redshifts, as is expected \citep{mortlock13}. Both of these types of systems were labeled as peculiar (morphology consistent with ongoing or post-interaction) in \cite{mortlock13} and we will keep the same terminology here when using their observed fractions.  Consequently, if the fraction of peculiar systems is expected to evolve with redshift, the mean value of the $\qfir$ parameter will also then be expected to change with redshift. In the most general way it will evolve as
\beqar
\label{eq:firz}
\nonumber
\bar{q}_{\rm FIR}(z) &=& \frac{\sum_i^N q_{i,\rm FIR} }{N} = \frac{N_{\rm alld} q_{\rm FIR} +  N_{\rm p} q_{\rm FIR,p}}{N}  \\
&=& q_{\rm FIR} (z) - \delta \qfir(z) \frac{N_{\rm p}}{N}
\eeqar
where $q_{\rm FIR}$ is the mean FIR-radio correlation parameter in the non-interacting, star-forming disk galaxies, $q_{\rm FIR,p}$  is the mean parameter for peculiar galaxies, and both can in general be functions of redshift. Difference between the two values is given as $\delta \qfir = q_{\rm FIR} - q_{\rm FIR,p} = \log (1+ S_{1.4} / S_{1.4, d})$, and where $N=N_{\rm alld}+N_{\rm p}$, $N_{\rm alld}$ and $N_{\rm p}$ are numbers of total, all disk (disk + perturbed disk) and peculiar galaxies in the sample, respectively. Let us assume that the fraction of peculiar galaxies is a known function of redshift. In our analysis we adopt these fractions from \cite{mortlock13}: $f_{\rm alld}$ is the fraction of all disk type galaxies with respect to the total number of galaxies in the sample given as $f_{\rm alld}= 3.88 (1+z)^{-3.30}$, and  $f_{\rm p}$ is the fraction of all peculiar (irregular + interacting) types of galaxies given as $f_{\rm p}=0.06(1+z)^{1.58}$ (these were obtained by fitting the results presented their Figure 5). If we look at all peculiar systems (denote these as $P$ models) we then find 
\beqar
\nonumber
\frac{N_{\rm p}}{N}  &=& \frac{N_{\rm p}}{N_{\rm p} + N_{\rm alld}}= \frac{1}{1+N_{\rm alld}/N_{\rm p}}=\frac{1}{1+f_{\rm alld}/f_{\rm p}}  \\
&=& \frac{1}{1+64.7(1+z)^{-4.88}}
\eeqar
Though $q_{\rm FIR}$ might be expected to change in interacting galaxies as the interaction proceeds due to particle acceleration in processes other than star-formation \citep{murphy13,dp2015}, it is unclear what would be the underlying cause for $q_{\rm FIR}$ deviation in irregular (post-interacting) systems. 
On the other hand, if the dispersion and evolution of the $\qfir$ is only due to interacting systems, we construct another set of models (denote these as $I$ models), where in that case $N_{\rm  p} \equiv N_{\rm int}$ and $N_{\rm p}/N$ is substituted with
\beqar
\nonumber
\frac{N_{\rm int}}{N}  &=& \frac{N_{\rm int}}{N_{\rm int} + N_{\rm d}}= \frac{1}{1+N_{\rm d}/N_{\rm int}}=\frac{1}{1+f_{\rm d}/f_{\rm int}} \\
&=&\frac{1}{1+3.6(1+z)^{-1.61}}
\eeqar
where $N_{\rm d}$ is the number of unperturbed disk galaxies and $f_{\rm d}=0.90(1+z)^{-2.80}$ and $f_{\rm int}=0.25(1+z)^{-1.19}$ are fractions of pure disk and interacting galaxies respectively, as taken from \cite{mortlock13}.
Besides having two classes of models depending on which type of systems we assume are the source of this discrepancy,  all peculiar systems or only interacting systems, we also need to consider how large is this discrepancy, that is, how much does $\qfir$ in peculiar or interacting systems deviate from the nominal value in normal, disk, star-forming systems. We will explore few possibilities and model them separately: model (a) where $\delta \qfir$ is a constant free parameter, model (b) where we assume that $q_{\rm FIR,p}$ intrinsically evolves according to findings for peculiar galaxies presented in Section \ref{sec:qmorphology} as $q_{\rm FIR,p} =4.14 (1+z)^{-0.62}$, models (c) \& (d) where we assume some underlying mechanism as the cause of $\qfir$  deviation in irregular and interacting galaxies. Widely accepted wisdom says that the FIR-radio correlation follows naturally from the process of star-formation that is traced by both infrared and radio emission. Thus any deviation from this correlation has to arise from some additional process that affects one or the other in a greater way. For example, enhanced FIR emission would be expected if there was an additional, important dust-heating process at place, such as shock heating due to shocks that arise from galactic interactions \citep{dp2015}. On the other hand, enhanced synchrotron emission could result from magnetic field being amplified in galactic interactions, since \cite{drzazga2011} have found that even a weak interaction results in the amplification of magnetic field up to 2 times. In that case, if the magnetic field is on average amplified to $B \sim 2 B_0$ during interaction from its initial value $B_0$ (neglecting here the evolution of magnetic field), we would have a constant $\delta \qfir = q_{\rm FIR,0} -  q_{\rm FIR,p} = q_{\rm FIR,0} - \log F_{\rm IR,p} / S_{\nu,p} = q_{\rm FIR,0} - \log F_{\rm IR,0} +  \log \left[ S_{\nu,0} (1 +   \log S_{\nu,p}/ S_{\nu,0} ) \right] = \log (1 +   S_{\nu,p}/ S_{\nu,0} ) = \log (1+(B/B_0)^2) = 1$. Here $F_{\rm IR,p} $ and $S_{\nu,p}$ refer to the infrared and radio emission in peculiar (irregular and/or interacting) galaxies respectively. This will be labeled as  model (c). Another way to have enhanced synchrotron emission would be if there was a new  population of cosmic rays, in addition to standard galactic cosmic rays, which could be accelerated in tidal shocks, or injected in some burst event related to gas accretion in galactic center, following galactic interaction. This could be related to the available gas mass fraction\footnote{For example, occurrence of large-scale tidal-shocks in interacting galaxies could lead to acceleration of a cosmic-ray population that would be competing with galactic cosmic rays \citep{pub13}.} and thus, in model (d) we will consider
$\delta \qfir = \log (1 +   S_{\nu,p}/ S_{\nu,0} ) = \log \left[ 1 +   f_{\rm gas} (z) \right]= \log \left[ 1 +   0.045 (1+z)^{1.31} \right] $, where we have adopted the gas fraction $f_{\rm gas}$ from \cite{santini14} (Figure 12 of their paper).

Finally we have to consider that the FIR-radio correlation in the non-interacting disk galaxies can itself evolve. We thus construct a set of models that allow for evolution of $q_{\rm FIR,0}$, due to the increasing Inverse Compton losses with increasing redshift, by fitting the results presented in \cite{murphy2009} for magnetic field strengths of $B=10 \,\mu$Gauss and $B=20 \, \mu$Gauss respectively as $q_{\rm FIR,B10}=2.01+0.66(1+z)-0.086(1+z)^2+0.004(1+z)^3$ and $q_{\rm FIR,B20}=2.13+0.42(1+z)-0.04(1+z)^2+0.0009(1+z)^3$. These models will be labeled as $B$ models. For example, a model labeled as IB1020 allows for the evolution of $\qfir$ due to the increasing Inverse Compton losses, where for normal, unperturbed and non-interacting galaxy, a magnetic field of $10 \, \mu$Gauss is assumed, while for interaction phase we assume that magnetic field doubles, thus we model its $\qfir$ evolution due to increasing Inverse Compton losses with results from \cite{murphy2009} for $20 \, \mu$Gauss. 

\begin{figure}[H]
\centering
\includegraphics[width=0.65\textwidth]{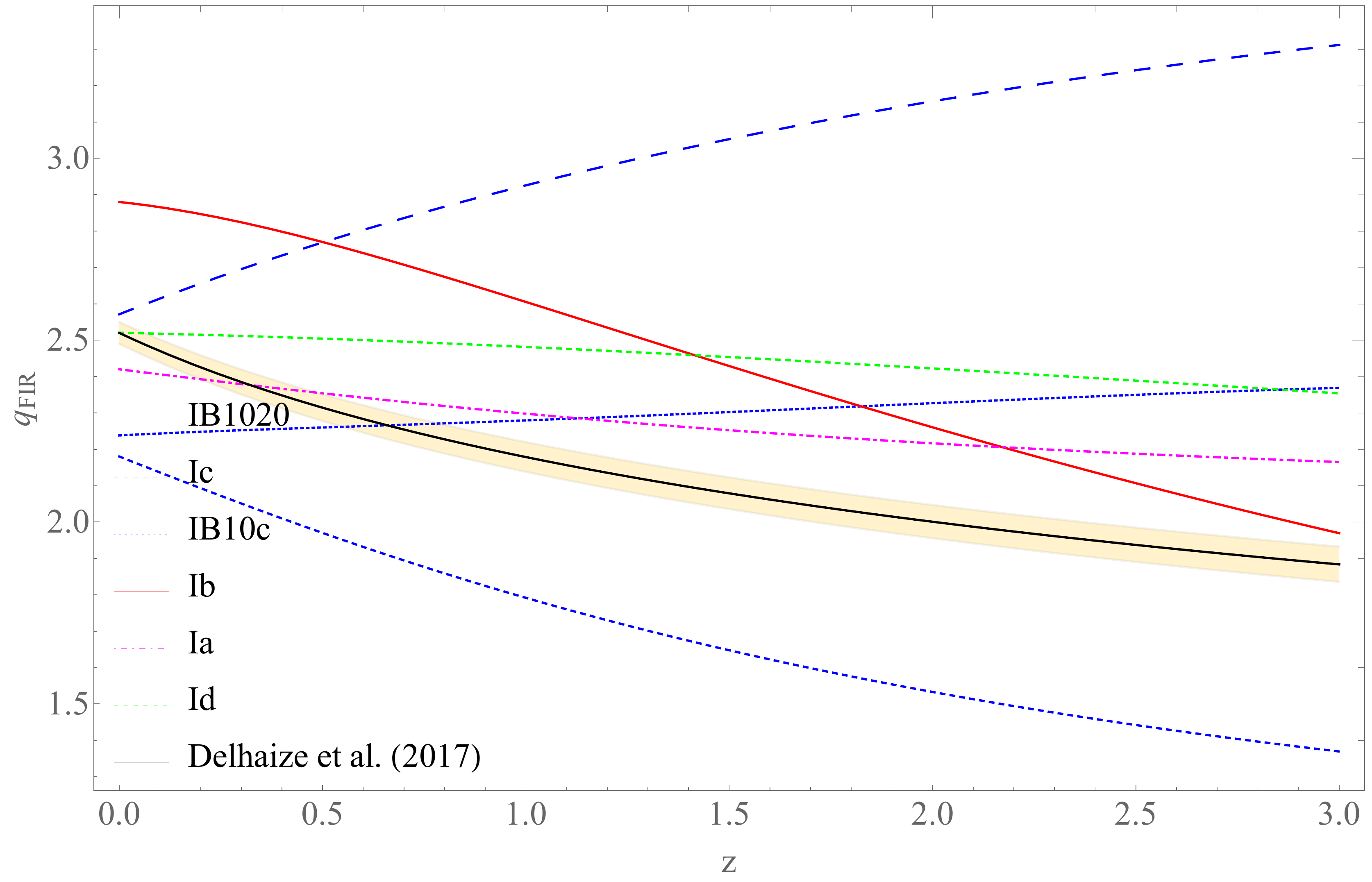}
\caption{\label{fig:modelsI} 
Models of $\bar{q}_{\rm FIR}$ evolution versus redshift compared to evolution found in \citet{delhaize2017} (solid, black curve). Models assume that the presence of the interacting galaxies in the sample is driving this evolution ("I models"). Different models are represented with following curves  (see Table \ref{table:models}): Large, dashed blue curve - model IB1020; small, dashed blue curve - model Ic; dotted, blue curve - model IB10c; solid, red curve - model Ib; dot-dashed, magenta curve - model Ia; small, dashed, green curve - model Id.}
\end{figure}

\begin{figure}[H]
\centering
\includegraphics[width=0.65\textwidth]{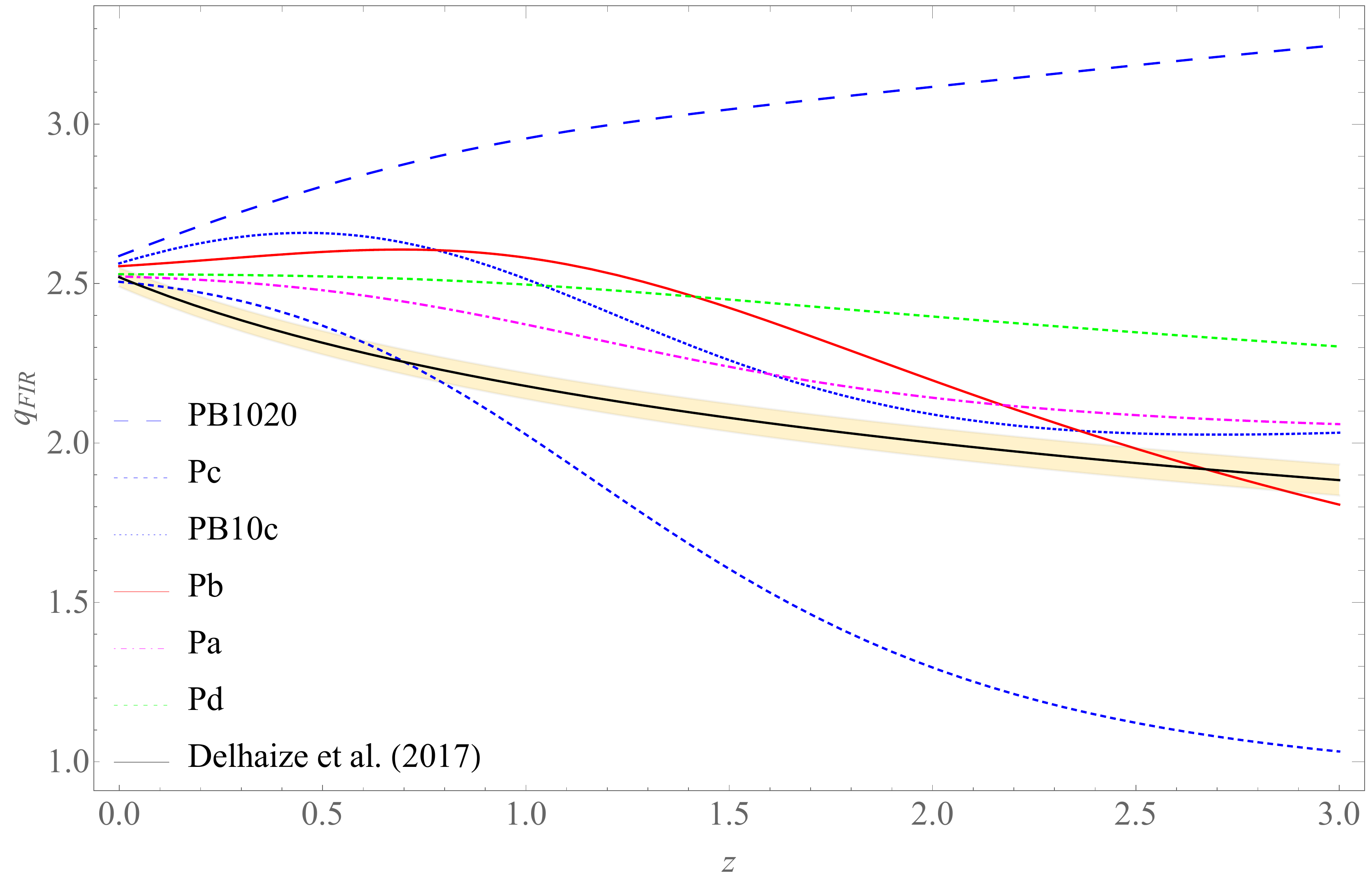} 
\caption{\label{fig:modelsP} 
 Same as on the Fig. \ref{fig:modelsI} but here assume that presence of peculiar (interacting and irregular) galaxies in the sample is driving this evolution ("P models"). Different models are represented with following curves  (see Table \ref{table:models}): Large, dashed blue - model PB1020; small, dashed blue curve - model Pc; dotted, blue curve - model PB10c; solid, red curve - model Pb; dot-dashed, magenta curve - model Pa; small, dashed, green curve - model Pd.}
\end{figure}

A summary of all models constructed is given in the Table \ref{table:models} and the corresponding results are plotted on Figures \ref{fig:modelsI} and \ref{fig:modelsP}. Most models presented in this work, which are based on the assumption that galactic interactions can affect the FIR-radio correlation, exhibit a decrease with redshift in the FIR-radio ratio. As pointed out earlier, the current wisdom that comes from theory is that either the FIR-radio correlation is non-evolving or that it should be increasing with redshift. Here, for the first time, we demonstrate that there are effects that could result in a decreasing trend, as was found by  \cite{delhaize2017}. The model that possibly best matches the behavior observed in \cite{delhaize2017} (presented with solid, black line) is the model Ia, which is plotted with dash-dotted, magenta curve. We caution the reader that model Ib (solid, red curve on Fig.\ref{fig:modelsI}), based on the trend found for peculiar galaxies, as described in the next section, and applied on the interacting systems, is also the one with the largest uncertainty due to a very small number statistics. Figures \ref{fig:modelsI} and \ref{fig:modelsP} represent models that assume sources of different type to be the main couse of this evolution - all peculiar galaxies in one case (irregular and interacting) presented on the right panel, and only interacting galaxies presented on the left panel. We see that some  models show similar behavior, especially in the case of Peculiar modes, which means that it will be more difficult to discriminate between them if there is a large data scattering. On the other hand, models that include effects of magnetic fields (c, B10 and B1020) show large divergence at higher redshifts, thus making it much easier to discriminate between them.
It is important to point out that presented models do not include uncertainties which can, in some cases, be quite large and difficult to estimate. The largest uncertainties are coming from the unknown level of particle acceleration in large-scale shocks arising from galactic interactions, unknown evolution of galactic magnetic fields and their amplification in galactic interactions, increasing difficulty to resolve galactic interactions and stages at larger redshifts. Furthermore, though the quoted and plotted error in the \cite{delhaize2017} fit is statistical and small, there is still significant scatter present in the data, and models presented here can still provide explanation for some level of scatter for both high and low $q_{\rm FIR}$ values originating from different interaction stages.

\begin{table}[H]
\caption{Different models of $\qfir$ evolution. The first column is the model label, the second column is the value of $\bar{q}_{\rm FIR}(z)$ as in Eq. (\ref{eq:firz}), the third column are different choices of the $\delta \qfir$. Models in this table have labels starting with $I$ (interacting) and correspond to curves presented on Fig. \ref{fig:modelsI}. Equivalent set of models labeled as "$P$ models" for all peculiar galaxies, which instead of $N_{\rm int}/N$ have $N_{\rm p}/N$ are shown on Fig. \ref{fig:modelsP}.}

\label{table:models}
\hfill \break
\centering
\begin{tabular}{lcc}
\hline
Model label & $\bar{q}_{\rm FIR}(z)$ & $\delta \qfir$ \\
\hline
Ia  & $q_{\rm FIR,0}  - \delta \qfir (N_{\rm int}/N) $ & $0.2q_{\rm FIR,0}$ \\
Ib  & $q_{\rm FIR,0}  - \delta \qfir  (N_{\rm int}/N) $ & $q_{\rm FIR,0}-4.14(1+z)^{-0.62} $ \\
Ic  & $q_{\rm FIR,0}  - \delta \qfir  (N_{\rm int}/N)  $ & $\log [1+ (B/B_0)^2]$ \\
Id  & $q_{\rm FIR,0}  - \delta \qfir  (N_{\rm int}/N)  $ & $\log [1+ 0.045(1+z)^{1.31}]$ \\
IcB10& $q_{\rm FIR,B10} (z)  - \delta \qfir  (N_{\rm int}/N) $ & $\log [1+ (B/B_0)^2]$ \\
IB1020 & $q_{\rm FIR,B10} (z) - \delta \qfir  (N_{\rm int}/N) $ & $q_{\rm FIR,B10} (z) - q_{\rm FIR,B20} (z) $ \\
\hline
\end{tabular}
\end{table}

\section{Data}

To test models presented in the previous section and observed decrease in the $\qfir$ as found in \cite{delhaize2017}, we now turn to analysis of available data. From the data set that was  already published in \cite{miettinen2017b} we select 34 SMG galaxies, 11 of which were identified as peculiar (irregular and/or interacting), and 23 of which were identified as disk galaxies. We summarize their properties in the Table \ref{table:sample}. Galaxies suspected of harboring an active galactic nuclei were not included in the sample. We have only selected galaxies that had observed fluxes on at least 4 wavelengths in the FIR (from $24\rm \mu m$ to $250\rm \mu m$), in order to be able to fit the SED curves in this range.
The chosen SMGs were discovered by the $\lambda_{\rm obs} = 1.1 \, \rm mm $ blank-field continuum survey over an area of 0.72 $\rm deg^2$ or $37.5\%$ of the full $2\rm \, deg^2$ COSMOS field, conducted with the AzTEC bolometer array on the 10m Atacama Submillimeter Telescope Experiment \citep[ASTE]{ezawa04}. The angular resolution of these observations was $34 \rm''$ FWHM.

\begin{table}[H]

\caption{Characteristic of 34 SMGs. The meaning of the columns is: (1): SMG name; (2) and (3): celestial peak position (equinox J2000.0) taken from \citet{miettinen2017b}; (4): redshift \citep[for more information see][]{miettinen2017b}; (5): radio spectral index determined from 1.4GHz and 3GHz observed-frame frequencies \citep[table C.1.] []{miettinen2017b}  (6): morphology type of galaxies, where disk and pec stands for disk and peculiar (irregular and/or interacting) galaxies, respectively \citep[Table 4 in ][]{miettinen2017b}; (7): rest-frame far-infrared flux density in the range of $42 \rm \mu m$ to $122 \rm \mu m$, calculated from observations at 4 wavelengths ($24 \rm \mu m, 100 \rm \mu m, 160 \rm \mu m$ and $250 \rm \mu m$) taken from Herschel continuum observations \citep{pilbratt10, miettinen2017a}; (8): observed-frame radio flux density at $1.4 \rm GHz$ taken from VLA COSMOS survey \citep{schinnerer10,aretxaga11,miettinen2017a}; (9): parameter $q_{\rm FIR}$ determined from Eq. (\ref{eq:q}).}

\label{table:sample}
\hfill \break
{\footnotesize
\begin{tabular}{lcccccccc}
\hline
Source & $\delta$ & $a$ & $z$ & $ \alpha _{1.4 \rm GHz}^{3 \rm GHz}$ & Type & $ \rm FIR$ & $S_{1.4 \rm GHz}$ & $q_{\rm FIR}$ \\
 &[$^\circ$:':''] & [h:m:s] & & & & [$\rm mJy$] & [$\mu \rm Jy$] & \\
\hline
C2a & +02 34 41.05 & 09 59 59.33 & 3.18 & $-0.95 \pm 0.32$  & disk & $25.80 \pm 12.68$ & $102 \pm 13$ & $2.43 \pm 0.28$\\
C6a & +02 20 17.31 & 10 00 56.95 & 2.49 & $-0.76 \pm 0.39$ & disk & $13.39 \pm 12.08$ & $48 \pm 12$ & $2.58 \pm 0.46$\\
C13a & +02 14 08.43 & 09 58 37.97 & 2.01 & $-0.73 \pm 0.13$ & disk & $9.66 \pm 4.25$ & $144.2 \pm 13.3$ & $1.97 \pm 0.21$\\
C16b & +02 16 45.95 & 09 58 54.19 & 2.39 & $-0.99 \pm 0.40$ & disk & $8.61 \pm 6.50$ & $82.1 \pm 13.8$ & $2.02 \pm 0.39$\\
C18 & +02 43 53.27 & 10 00 35.30 & 3.15 & $-0.82 \pm 0.34$ & pec & $16.45 \pm 15.88$ & $98 \pm 16$ & $2.34 \pm 0.47$\\
C22a & +02 40 10.90 & 10 00 08.94 & 1.60 & $-0.37 \pm 0.27$ & pec & $15.10 \pm 13.65$ & $132 \pm 26$ & $3.32 \pm 0.15$\\
C22b & +02 40 09.52 & 10 00 08.90 & 1.60 & $-0.94 \pm 0.30$ & pec & $6.13 \pm 13.65$ & $138 \pm 26$ & $1.67 \pm 0.20$\\
C23 & +02 18 35.88 & 10 01 42.36 & 2.10 & $-0.73 \pm 0.22$ & disk & $10.77 \pm 10.64$ & $124.5 \pm 13.9$ & $2.02 \pm 0.45$\\
C24a & +02 22 24.42 & 10 00 10.36 & 2.01 & $-1.07 \pm 0.37$ & disk & $7.39 \pm 6.83$ & $99.6 \pm 12.6$ & $1.83 \pm 0.46$ \\
C25 & +01 56 43.57 & 10 01 21.95 & 2.51 & $-0.53 \pm 0.54$ & disk & $13.03 \pm 11.38$ & $70.2 \pm 12.2$ & $2.52 \pm 0.45$\\
C28a & +02 13 01.64 & 09 58 49.28 & 2.32 & $-1.43 \pm 0.19$ & disk & $9.04 \pm 2.58$ & $382.0 \pm 52.1$ & $1.17 \pm 0.16$\\
C32 & +02 01 24.22 & 10 00 12.53 & 1.63 & $-0.28 \pm 0.47$ & disk & $12.50 \pm 3.11$ & $55.9 \pm 10.9$ & $2.73 \pm 0.28$\\
C33a & +02 31 40.77 & 10 00 27.14 & 2.30 & $-0.49 \pm 0.34$ & disk & $24.78 \pm 2.76$ & $67.3 \pm 11.7$ & $2.79 \pm 0.17$\\
C36 & +02 05 14.58 & 09 58 40.29 & 2.41 & $-0.99 \pm 0.21$ & pec & $17.79 \pm 2.45$ & $167.6 \pm 14.9$ & $2.03 \pm 0.13$\\
C43a & +02 02 01.53 & 10 00 03.12 & 2.01 & $-2.15 \pm 0.32$ & disk & $7.48 \pm 3.10$ & $190 \pm 47.9$ & $1.04 \pm 0.26$\\
C46 & +02 35 18.36 & 10 01 14.71 & 1.06 & $-1.01 \pm 0.16$ & disk & $7.44 \pm 8.59$ & $122 \pm 12.4$ & $1.78 \pm 0.51$\\
C47 & +02 01 13.25 & 09 59 40.87 & 2.05 & $-1.49 \pm 0.12$ & pec & $7.17 \pm 2.60$ & $330 \pm 32.4$ & $1.10 \pm 0.17$\\
C52 & +02 21 00.93 & 10 01 56.57 & 1.15 & $-0.92 \pm 0.36$  & pec & $15.74 \pm 9.72$ & $119.1 \pm 15.4$ & $2.15 \pm 0.30$\\
C59 & +02 37 16.76 & 10 00 30.14 & 1.28 & $-1.19 \pm 0.14$ & pec & $16.84 \pm 8.60$ & $161.1 \pm 14.6$ & $1.95 \pm 0.23$\\
C65 & +02 21 44.91 & 09 59 42.94 & 1.80 & $-0.91 \pm 0.20$ & pec & $15.68 \pm 7.93$ & $153.4 \pm 12.1$ & $2.05 \pm 0.24$\\
C66 & +02 26 33.98 & 10 01 04.64 & 2.01 & $-0.66 \pm 0.32$ & pec & $12.55 \pm 10.22$ & $86 \pm 11$ & $2.33 \pm 0.39$\\
C67 & +02 09 44.67 & 10 01 19.53 & 2.93 & $-0.92 \pm 0.26$ & disk & $6.95 \pm 5.32$ & $70 \pm 12.3$ & $2.03 \pm 0.36$\\
C72 & +02 04 57.67 & 10 01 58.99 & 1.72 & $-0.84 \pm 0.32$ & disk & $10.75 \pm 4.34$ & $95.5 \pm 13.7$ & $2.12 \pm 0.23$\\
C77b & +01 57 59.20 & 09 59 35.30 & 3.06 & $-1.24 \pm 0.30$ & disk & $8.96 \pm 10.23$ & $69.1 \pm 10.7$ & $1.99 \pm 0.52$\\
C84b & +01 55 01.49 & 09 59 42.58 & 1.96 & $-0.65 \pm 0.41$ & pec & $8.96 \pm 11.09$ & $82.4 \pm 10.9$ & $2.20 \pm 0.57$\\
C93 & +02 11 38.77 & 10 01 31.88 & 1.63 & $-0.57 \pm 0.26$ & disk & $7.38 \pm 5.35$ & $60.1 \pm 11.1$ & $2.30 \pm 0.35$\\
C97b & +02 19 42.84 & 10 02 14.50 & 2.01 & $-1.06 \pm 0.76$ & disk & $7.46 \pm 3.85 $ & $55.1 \pm 15.9$ & $2.10 \pm 0.46$\\
C98 & +02 05 19.03 & 10 00 43.18 & 1.82 & $-0.83 \pm 0.40$ & disk & $14.97 \pm 8.09$ & $77.8 \pm 13.7$ & $2.37 \pm 0.31$\\
C109 & +02 28 40.89 & 10 01 11.56 & 2.20 & $-1.43 \pm 0.75$ & disk & $13.67 \pm 6.35$ & $59.1 \pm 10.6$ & $2.10 \pm 0.51$\\
C111 & +02 12 43.97 & 09 59 29.23 & 2.10 & $-0.81 \pm 0.27$ & disk & $8.29 \pm 6.16$ & $66.7 \pm 11.6$ & $2.19 \pm 0.36$\\
C112 & +01 53 14.06 & 10 00 11.03 & 1.89 & $-0.82 \pm 0.23$ & pec & $9.53 \pm 10.23$ & $121.8 \pm 11.5$ & $1.98 \pm 0.48$\\
C113 & +02 29 60.00 & 09 59 14.40 & 2.09 & $-0.63 \pm 0.11$ & disk & $31.25 \pm 48.98$ & $173 \pm 15.6$ & $2.41 \pm 0.68$\\
C116 & +02 03 46.43 & 10 01 09.85 & 2.20 & $-0.64 \pm 0.27$ & disk & $7.20 \pm 11.91$ & $58.9 \pm 10.8$ & $2.18 \pm 0.73$\\
C127 & +02 35 27.32 & 10 01 25.33 & 2.01 & $-0.98 \pm 0.15$ & disk & $14.77 \pm 9.77$ & $131.1 \pm 12.5$ & $2.06 \pm 0.30$\\

\hline
\end{tabular}}
\end{table}

The morphology of these sources has already been determined and published in the Tasca and Cassata Catalogues \citep{tasca09, cassata07}. The morphological classification for our selected data set was summarised in the Table 4 of \cite{miettinen2017b}, where the Authors adopt terminology {\it disk galaxy} for spiral, star-forming galaxies and terminology {\it irregular} for single, irregular galaxies. We have adopted the same terminology and took 11 irregular galaxies from their Table 4, where the flux density was available on four wavelengths (from $24 \rm \mu m$ to $250 \rm \mu m$). We have used available observed flux densities to fit the SED curve in the FIR range. In the selected sample of SMGs, sources AzTEC/C22a and AzTEC/C22b are good candidates for interacting, Taffy-like systems \citep{miettinen2017b}, so we also include them in our peculiar subsample. The ALMA 1.3 mm image of the AzTEC/C22 shows the presence of two dust-emitting cores which are separated by $13.8 \rm \, kpc$, and associated with $3 \rm GHz$ radio-emitting bridge that connects them. A well separated binary structure of AzTEC/C22a and AzTEC/C22b points out to an early-stage SMG-SMG or an incomplete merger, and one portion of the radio emission from these galaxies is probably coming from a magnetized medium between them \citep{miettinen2017b}, and possibly from cosmic rays accelerated in tidal shocks in their discs \citep{dp2015}. Looking at each of their morphologies separately, these systems were identified as disk galaxies. However, because we want to explore if interactions, ongoing or recent, could affect the evolution of the FIR-radio correlation, and in order to be consistent with models presented in the previous section, we will label these galaxies and other similar systems as peculiar to describe both interacting and irregular systems.

The FIR ($24\rm \mu m, 100\rm \mu m, 160\rm \mu m,
$ and $ 250\rm \mu m$) to submm ($350\rm \mu m $) data for our selected sample were taken from the Herschel  continuum observations \citep{pilbratt10, miettinen2017a}, and obtained as a part of the Photodetector Array Camera and Spectrometer (PACS) Evolutionary Probe \citep[PEP;][]{lutz11} and the Herschel Multi-tiered Extragalactic Survey \citep[HerMES;][]{oliver12} programs.
The total, rest frame, FIR flux was calculated for each galaxy separately from their SED curve from the observed-frame infrared flux densities at 4 wavelengths ($24\rm \mu m, 100\rm \mu m$, $160\rm \mu m$ and $250\rm \mu m$), and fitted by a second-degree polynomial. The total $\rm FIR$ flux density was then obtained by integrating the SED curve in the rest frame in the range of $42\rm \mu m$ to $122\rm \mu m$, which was than used to calculate the parameter $q_{\rm FIR}$. Errors were propagated and combined in quadrature.

In this paper we have used the observed-frame radio flux density at $1.4 \rm GHz$ from the VLA (Very Large Array)-COSMOS survey \citep{schinnerer10, aretxaga11, miettinen2017a}. Using spectral available indices (see table \ref{table:sample}), we have recalculated the rest-frame flux densities \citep{magnelli15} at $1.4 \rm GHz$, and used them for determining the $q_{\rm FIR}$ parameter.  

\subsection{FIR-radio correlation vs. Morphology}
\label{sec:qmorphology}

Our chosen sample of 34  SMGs (peculiar + disk galaxies) has  the FIR-radio correlation parameter with the mean value of  $q_{\rm FIR}=2.11\pm 0.08$ (quoted error is statistical). When comparing this to the nominal value of $q_{\rm FIR,0}=2.34\pm 0.01$ \citep{yun01} that was derived from a much larger, low-redshift sample, we see that high-redshift SMGs show significantly lower mean FIR-radio correlation parameter. This supports the findings of \citet{delhaize2017} where  $q_{\rm FIR}$ decreases with redshift. 

To examine the dependence of the  $q_{\rm FIR}$ parameter evolution on morphology type, we divide the sample into two sub-samples, according to their morphology, and analyze them separately. Looking first into the mean values we find that a sub-sample of 11 peculiar galaxies has $q_{\rm FIR}=2.10\pm 0.15$ while the sub-sample of 23 disk galaxies shows $q_{\rm FIR}=2.12\pm 0.09$. Both of these sub-samples show consistently lower mean values than the low-redshift result of $q_{\rm FIR,0}=2.34\pm 0.01$ \citep{yun01}, with the lowest being that for peculiar galaxies but only slightly and with the largest uncertainty. 
To further compare the evolution of these samples, we try to fit the dependence of the $q_{\rm FIR}$ on $z$ in both cases by a power-law function  $q_{\rm FIR}(z)= a(1+z)^b$, where $a$ is a constant, and $b$ is the degree coefficient. The fitting function of this type was chosen to be be easily compared to the results of \cite{delhaize2017}.

\begin{figure}[H]
\centering
 \begin{tabular}{cc}
\includegraphics[width=0.45\textwidth]{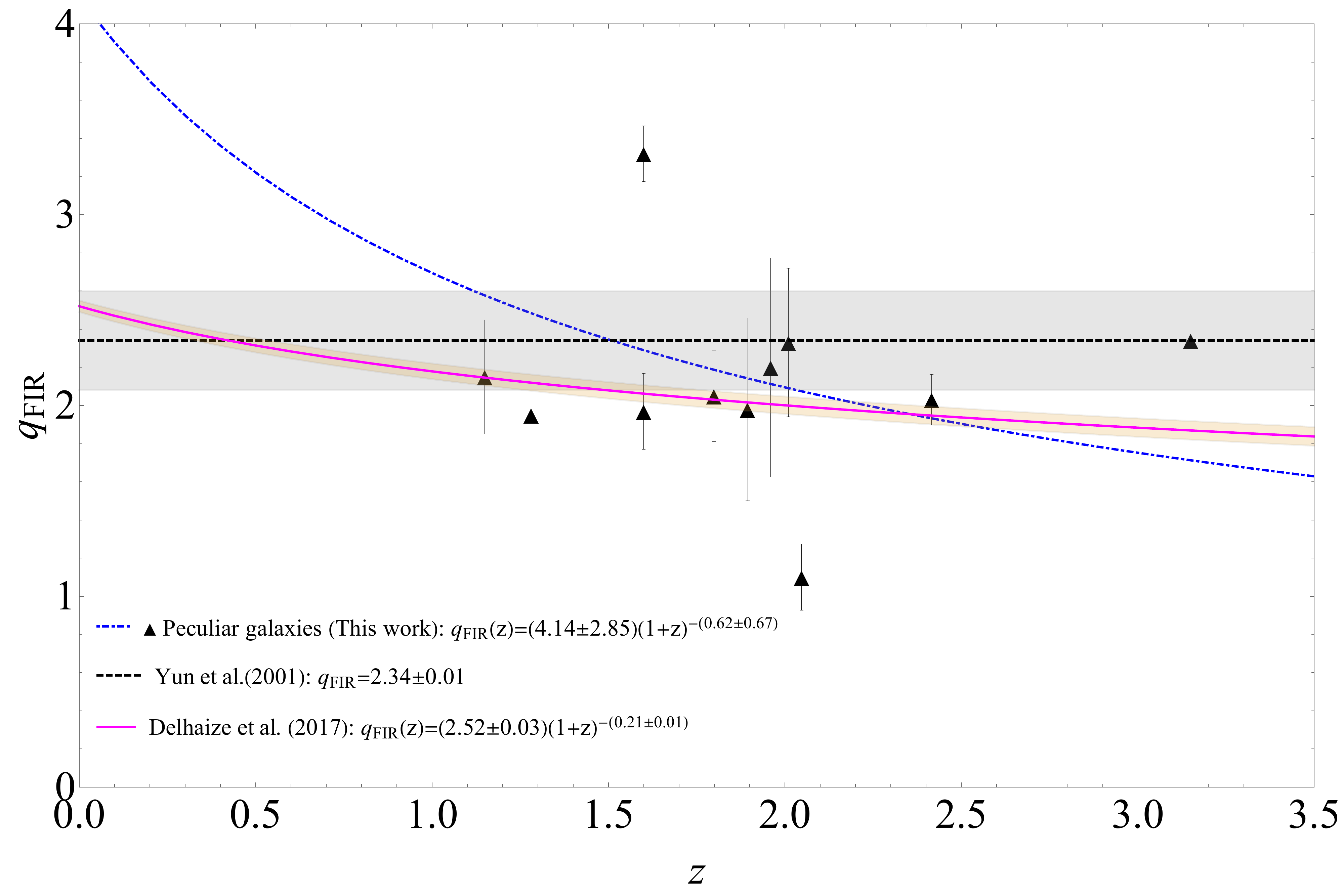} & 

\includegraphics[width=0.45\textwidth]{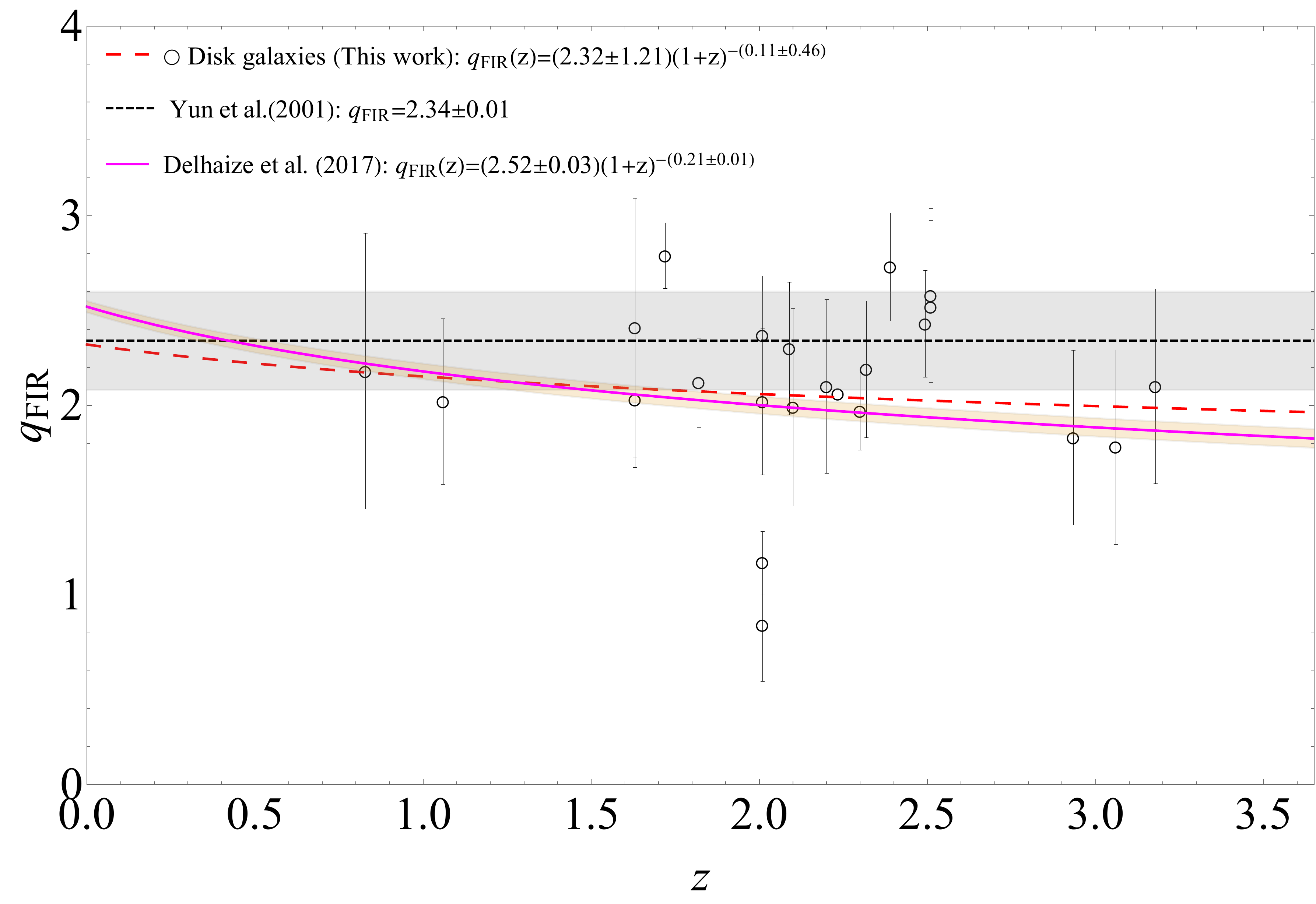} \\
\end{tabular}
\caption{$q_{\rm FIR}$ as a function of redshift. Left: data for 11 peculiar galaxies (black triangles) were fitted with the blue dash-dotted line. Right: data for 23 disk SMGs (black empty circles) were fitted with red dashed line. Solid pink curve represent result from \citep{delhaize2017}. Black dashed line is nominal value for $q=2.34 $ with gray-shaded region representing its standard deviation $\sigma =0.26$ \citep{yun01}.}
\label{tab:a}
\end{figure}

Figure \ref{tab:a} shows $q_{\rm FIR}$ as a function of redshift for 11 peculiar (left) and 23 disk galaxies (right). The blue, dash-dotted line and red, dashed line, correspond to the power-law fits $q_{\rm FIR}=(4.14\pm 2.85)(1+z)^{-(0.62\pm 0.67)}$ and $q_{\rm FIR}=(2.32\pm 1.21)(1+z)^{(-0.11\pm 0.46)}$, for peculiar and disk galaxies respectively. Though both samples hint possible evolution, with apparently stronger evolution in peculiar galaxies, both samples are also consistent with no evolution, due to large uncertainty and low-number statistics. Note that the choice of power-law fitting in the form of $a(1+z)^b$, together with the lack of data below redshift $\sim 1$, allows for a more steeper power-law index $b$ and a large normalization constant $a$, such that, at redshift $z=0$, fits go far above the nominal value $q=2.34$ \citep{yun01}. Removing outliers does not result in much change. However, if for example, we perform a linear regression, we get similar decreasing trends, but with more realistic values at zero redshift, specifically, linear fits for peculiar and disk galaxy samples yield $q_{\rm FIR} (z)=(3.05\pm 0.96) - (0.47 \pm 0.49)z$ and  $q_{\rm FIR} (z)=(2.16\pm 0.68) - (0.05 \pm 0.32)z$ respectively.
Lastly, it is important to note that when performing Kolmogorov-Smirnov test to compare the two subsamples, one finds that they are consistent with originating from the same distribution, which could also possibly be due to  small sample sizes and large uncertainties.

\begin{figure}[H]
\centering
\includegraphics[width=0.75\textwidth]{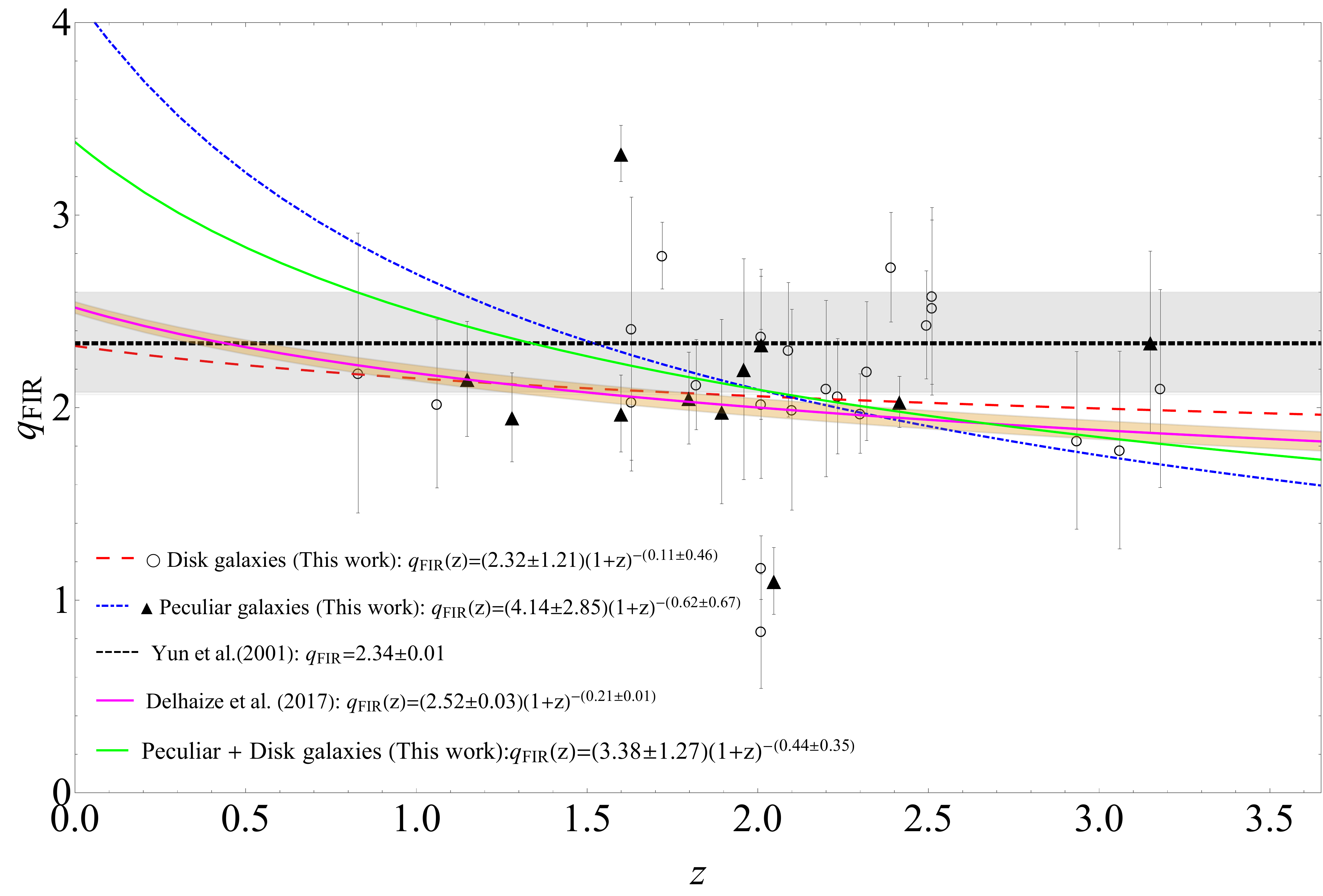}
\caption{\label{fig:both1} 
$q_{\rm FIR}$ versus redshift for 34 SMGs. Black empty circles represent disk galaxies while black triangles are peculiar galaxies. Red dashed and blue dash-dotted lines represent power-law fits for disk and peculiar galaxies, respectively. The green, solid, curve shows power-law fit for the whole sample of 34 SMGs (disk + peculiar) galaxies. Pink, solid, curve is the same power-law fit found in \citep{delhaize2017}. Black dashed line is nominal value for $q=2.34$, with gray-shaded region representing its standard deviation $\sigma =0.26$ \citep{yun01}.}
\end{figure}

Finally, we analyzed the entire sample together (disk + peculiar galaxies), which is shown on Figure \ref{fig:both1}, with green, solid curve, and find that the entire sample does show evolution with redshift with the power-low shape as $q_{\rm FIR}(z) = (3.38 \pm 1.27)(1 + z)^{(-0.44 \pm 0.35)}$.  The same plot also shows all curves, for disk (red, dashed line) and peculiar (blue, dash-dotted line) galaxies alone, and compares everything to results found in \cite{delhaize2017} that are presented with pink, (lower) solid curve. 
The black triangles represent 11 peculiar galaxies while black empty circles are 23 disk galaxies. 

\section{Conclusion}

The far-infrared radio correlation is a powerful tool used, among other things, for determining star-formation rates \citep{condon92}. Its power relies on the assumption that it remains stable across a large range of redshifts, though there is a significant scatter present. However, recent observation of star-forming galaxies in the COSMOS survey have revealed a decreasing trend towards higher redshifts and evolution in the form $q_{\rm FIR}=(2.52 \pm 0.03)(1+z)^{-0.21 \pm 0.01}$  \citep{delhaize2017}. 

In order to try to explain the observed trend, and inspired by claims that galactic interactions might result in cosmic-ray acceleration and affect the FIR-radio correlation \citep{lisenfeld10,murphy13,dp2015},  we have  modeled the expected evolution of the FIR-radio correlation ratio parameter $\qfir$, assuming that the underlying cause of evolution are galactic interactions. Our main assumption was that the sample of star-forming galaxies from which correlation is determined, presents a mix of star-forming galaxies of different morphologies - disk galaxies and galaxies that reflect current or past interactions. The evolution of the $\qfir$ then naturally follows from the evolving fraction of interacting and irregular galaxies, which we jointly call peculiar galaxies. Within this, we have explored several models, each of which assumes different mean $\qfir$ for interacting or irregular galaxies, with respect to the value in disk galaxies. Unlike other theoretical models that consider evolution of the FIR-radio correlation and find that if anything, an increasing trend is to be expected, most models presented in this paper based on interactions, rather show a decreasing trend with redshift, as was found by \cite{delhaize2017}. Some models are more diverging towards higher redshifts. This can be used to discriminate between models with more data available. However, given that uncertainties in the presented models are still large and difficult to reasonably estimate, while uncertainty of the evolution found in \cite{delhaize2017} is statistical and small, it is difficult, at this point, to completely exclude most presented models, except for the ones that show the increasing trend.

Furthermore, to test our models we have used the available sample of 34 sub-millimeter galaxies that have been observed and already morphologically classified \citep{miettinen2017a,miettinen2017b}. 
We have limited ourselves to this small sample of galaxies in order for make a consistent comparison with results of \cite{delhaize2017}. Obtaining a larger sample would require a reanalysis of observations and independent determination of morphologies, which was not intended at this point.
The sample was split into 23 pure disk galaxies and 11 peculiar (interacting and irregular) galaxies. Mean values of all three high-redshift samples (entire sample and two sub-samples) were found to be consistently and significantly below the low-redshift, large sample value of $q_{\rm FIR,0}=2.34 \pm 0.01$ \citep{yun01}. Furthermore we investigated the possible evolution with redshift within these samples, modeling it with a power-law shape in order for it to be easily comparable to findings of \cite{delhaize2017}.
Because both samples are very small, trends found when each sample is analyzed separately show evolution, but results are also, consistent with having no evolution with redshift:
$q_{\rm FIR}=(4.14\pm 2.85)(1+z)^{-(0.62\pm 0.67)}$ was found for peculiar galaxies, while $q_{\rm FIR} = (2.32 \pm 1.21)(1 + z)^{(-0.11 \pm 0.46)}$ for disk galaxies. However, when the entire sample is analyzed together a decreasing evolution was  found $q_{\rm FIR}= (3.38 \pm 1.27)(1 + z)^{(-0.44 \pm 0.35)}$. This is consistent with results found in \cite{delhaize2017} who found $q_{\rm FIR}$ to be decreasing with redshift when analyzed a large sample of star-forming galaxies. Furthermore, this indicates that one or both populations of different morphology are driving the observed decreasing trend, which will be revealed in larger samples.
While the analysis done on peculiar galaxies appears to show stronger evolution with redshift and has lower fractional uncertainty than in the case of pure disk galaxies, we caution the reader that both subsamples are consistent with originating from the same distribution when Kolmogorov-Smirnov test is performed.  We point out however, that this also might be due to small sample sizes, and that there is still a possibility that increasing fraction of irregular and interacting galaxies expected at larger redshifts might be driving the evolution observed in \cite{delhaize2017}. 

It is interesting to point out different evolution for different morphology found by \cite{molnar17}, who found a non-evolving IR/radio ratio for disk-dominated galaxies, but found a decreasing ratio for spheroid-dominated galaxies. This might be an indication of additional process contributing to the radio excess that is related to galactic interactions and/or active galactic nuclei activity. In the upcoming work we will analyze a much larger sample in order to test these results, where clear identification of systems with ongoing interaction and comparison of their IR/radio ratio to that in irregular and spheroidal systems will be important for discriminating between possible presence of active galactic nuclei or some new and additional process.

In conclusion -- we have presented a set of models that explore how changes in infrared and radio emission in interaction galaxies can reflect on the FIR-radio correlation. We have found that most such models would result in a evolving FIR-radio correlation, where its value would decrease towards higher redshift, consistent to what was found by \cite{delhaize2017}, and as opposed to current wisdom where if anything, an increase would be expected. We have also tested our models against a small set of available data, however due to small statistics, results are still inconclusive, as the data are consistent with no evolution but with very large uncertainties. On the other hand, when the entire sample was analyzed together, a decreasing trend was revealed. In order to better discriminate between models, analysis on a larger sample of pure disk, interacting and irregular galaxies is needed, which will be the focus of our future work.

\subsection*{Acknowledgments}

We are thankful to Darko Donevski for valuable discussion. We are also grateful to the Referee for useful comments which have helped improve this paper. The work of T.P. is supported in part by the Ministry of Science of the Republic of Serbia under project numbers 171002 and 176005.

\end{document}